\documentclass[a4,12pt]{article}
\usepackage{colortbl}
\usepackage{rotating}
\usepackage{makecell}
\usepackage{natbib}
\usepackage[dvipsnames]{xcolor}
\usepackage{booktabs}
\usepackage{booktabs}
\usepackage{afterpage}
\usepackage[normalem]{ulem}
\usepackage[top=4.65cm]{geometry}

\usepackage{hyperref}

\definecolor{britishracinggreen}{rgb}{0.0, 0.26, 0.15}
\definecolor{darkblue}{rgb}{0.0, 0.0, 0.55}
\definecolor{rred}{rgb}{25,0,255}
\definecolor{bbold}{rgb}{207,10,255}

\newcommand\B[1]{\cellcolor{yellow} \makecell{#1}}
\newcommand\A[1]{\cellcolor{green} \makecell{#1}}
\newcommand\C[1]{\cellcolor{orange} \makecell{#1}}
\newcommand\D[1]{\cellcolor{red} \makecell{#1}}
\newcommand\G[1]{\cellcolor{light-gray}\makecell[l]{#1}}

\title{On the use of artificial intelligence in financial regulations and the impact on financial stability\thanks{Corresponding author Jon Danielsson, J.Danielsson@lse.ac.uk. We thank Charles Goodhart, Gudmundur Kristjansson, Eva Micheler, Robert Macrae, Inaki Aldasoro, Leonardo Gambacorta, Robin Lumsdaine (discussant), Vatsala Shreeti and Bruno Tissot for valuable comments. Updated versions of this paper can be downloaded from from \url{modelsandrisk.org/appendix/AI}. We thank the Economic and Social Research Council (UK) [grant number ES/K002309/1] for their support. Any opinions and conclusions expressed herein are those of the authors and do not necessarily represent the views of the Bank of Canada.}
}

\author{Jon Danielsson\\\textit{London School of Economics}
\and Andreas Uthemann\\\textit{Bank of Canada}\\\textit{Systemic Risk Centre, London School of Economics}}

\date{June 2024\\{\scriptsize First version September 2023}  }

\begin{document}
\maketitle

\thispagestyle{empty}

\begin{abstract}
\noindent
Artificial intelligence (AI) can undermine financial stability because of malicious use, misinformation, misalignment, and the AI analytics market structure. The low frequency and uniqueness of financial crises, coupled with mutable and unclear objectives, frustrate machine learning. Even if the authorities prefer a conservative approach to AI adoption, it will likely become widely used by stealth, taking over increasingly high-level functions driven by significant cost efficiencies and superior performance. We propose six criteria for judging the suitability of AI.
\end{abstract}

\newpage

\section{Introduction}\label{introduction}

Artificial intelligence (AI)\footnote{\citet{Russel2019} is particularly useful for a general overview of AI. For a more technical review of the underlying ML techniques, see, e.g. \citet{pml2Book} and \citet{prince2023understanding}.} is transforming the financial system, improving its efficiency, robustness and impartiality. AI also creates new risks when known AI risk factors interact with established drivers of financial instability. Our objective in this work is to identify how AI can destabilise the financial system and the best ways for the authorities to respond to the challenges it raises. It is particularly beneficial to adopt the notion of AI as a rational maximising agent\footnote{It is one of \citet{RussellNorvig2010} notions in their classification of AI.} since that directly resonates with established channels of financial instability.

The financial system has a number of intrinsic vulnerabilities that are always at the centre of financial crises, such as data incompleteness, system responses, strategic complementarities and uncertainties. These can viciously interact with the internal logic of AI engines\footnote{\citet{weidinger2022taxonomy}, \citet{bengio2023managing} and \citet{shevlane2023model} discuss the societal risks arising from AI.} and how they are embedded in the financial system, with four channels of AI financial destabilisation of particular concern.

The \textit{malicious use channel} arises because the system is replete with highly resourced profit-maximising economic agents not too concerned about the social consequences of their activities. They seek to bypass controls and change the system in ways that benefit them while being difficult for other market participants and regulators to detect. Technology has always helped such agents, and they are already embracing AI. We expect the most common malicious use of AI will be by employees of financial institutions, careful to stay on the right side of the law, acting in a manner that is not only socially undesirable but even also against the interests of the institution employing them. AI will also facilitate illegal activities, such as rogue traders and criminals, as well as terrorists and nation-states aiming to create social disorder.

The \textit{misinformation channel} is due to the users of AI not understanding its limitations while increasingly becoming dependent on it. That is a particular concern for macro-prudential policies aiming to prevent and contain financial crises. In such applications, data is scarce and objectives unclear, so when data-driven algorithms are asked to extrapolate, they confidently advise on questions on which they have little or no pertinent data to train, a part of a broader phenomenon known as AI hallucination. 

The \textit{misalignment channel} emerges from difficulties in ensuring AI follows the objectives desired by its human operators. When faced with multiple and potentially conflicting objectives --- perhaps maximisation of profits, not breaking the spirit and the letter of the law, behaving ethically --- AI can end up prioritising objectives in an undesirable way. That is particularly likely when the engine is given abstract high level incentives while facing circumstances not in its training datasets. That can have several adverse consequences. One is the amplification of destabilising behaviour in times of crisis, which can happen for innocuous reasons, such as when financial institutions seek survival or withdraw liquidity, leading to destabilising dynamics such as fire sales, bank runs, and credit crunches. The problem of misalignment is particularly difficult when several AI engines interact, simultaneously learning from and reacting to each other, exploiting strategic complementarities that are prevalent in financial markets to create undesirable outcomes such as market manipulation and pricing cartels. It can be difficult to identify such behaviour ex-ante because AI will find it easy to evade oversight. 
 
The \textit{market structure channel} emanates from the business models of companies designing and running AI engines, which depend on three scarce resources: compute, human capital and data. That implies such firms enjoy increasing returns to scale, which can prevent entry, increase homogeneity and risk monoculture, even culminating in an oligopolistic market structure dominated by a few large vendors. All make the system more vulnerable to shocks while amplifying procyclicality, leading to more booms and busts. In other words, the oligopolistic nature of the AI analytic business increases systemic financial risk. Furthermore, those financial institutions with the best access to AI technology are pegged for winning, which not only leads to concentration and, hence, an oligopolistic market structure.

These four channels of AI-finance vulnerabilities present a difficult problem to the authorities, especially since most appear to prefer a slow, deliberative and conservative approach to AI, something not tenable. Several authorities have suggested policy responses, for example, \citet{LeitnerSinghKraaijZsamboki2024}, \citet{BISAI2024q} and \citet{Moufakkir2023}.

The authorities also face particular problems arising from its own use of AI. Along the way, they have to contend with two opposing forces. AI will increasingly be essential to the authorities working to keep the system stable, but it also aids the forces of instability. We suspect the second factor dominates, if only because the relative amount of compute needed for system-wide monitoring increases along with complexity. That is amplified by the intensification of the arms race between the authorities and the private sector that has always been with us. Both seek the same human resources, data and compute in a competitive market where the costs are much higher than the authorities are accustomed to.

Suppose the authorities also depend on the same AI engine for its analytics, which seems likely. Then, they may only be able to identify the resulting fragilities once a crisis has already happened. To help mitigate that, the engines must be able to explain how they arrive at their conclusions, in the process providing an assessment of statistical accuracy. That will not be sufficient, and the risk of monoculture will still be present. To overcome that, it becomes essential to run alternative analytics engines trained on different data and from several vendors. Here, it will be helpful if the authorities overcome their frequent reluctance to adopt consistent quantitative frameworks for measuring and reporting on the statistical accuracy of their data-based inputs and outputs. It is a concern that neither the competition nor the financial authorities have fully appreciated the potential for increased systemic risk due to oligopolistic AI technology in the recent wave of data vendor mergers.

Meanwhile, the public sector's use of AI in regulations raises tough questions on its own. Who is accountable when a regulator's AI makes decisions or provides crucial inputs for human decisions, and how can a regulated entity challenge decisions? The regulatory AI may not be able to explain its decision in terms that are intelligible to the subjects of regulatory actions, the AI's human owner or the judges involved in arbitration processes.

While all of this might suggest that the prudent way forward is to limit AI to strictly advisory roles, that will not be sustainable, as AI builds up trust and we come to depend on it, it will be the entity that gives advice. That means there may be no alternative to accepting its advice. In effect, AI becomes the decision-maker whether we want it or not.

We provide concrete suggestions to the authorities on how they can benefit from AI and the particular risks to watch out for along the way. We include six criteria for evaluating AI use and map those onto particular regulatory activities.

The organisation of the paper is as follows. After the introduction, we present the main channels of financial vulnerabilities in Section \ref{sec:conceptual} and follow that with a discussion on artificial intelligence and machine learning in Section \ref{sec:AI}. Section \ref{sec:GoodUse} is focused on the use of AI in the financial system, while Section \ref{sec:channels} contains the five channels for how AI can destabilise the financial system. We then apply those to the regulation of AI in Section \ref{sec:regulation}. Section \ref{conclusion} concludes.

\section{Channels of financial vulnerabilities}\label{sec:conceptual}

\begin{quote}
\textit{``Any observed statistical regularity will tend to collapse once pressure is placed upon it for control purposes.''}\vspace{-2mm} \flushright Charles Goodhart's (1974) Law.\nocite{Goodhart74}
\end{quote}

Any attempt at controlling the financial system is frustrated by the presence of highly resourced financial market participants operating in uncertain social environments that undergo frequent structural changes. The rules of the game are usually unclear, and the actors often do not know each others' objectives. Market participants can even change the rules to their advantage in a way that others only partially observe. They interact by competing, cooperating or colluding with each other, learning from interactions with the system, continually changing their beliefs and, hence, how they will react in similar situations in the future. This has always frustrated attempts at exercising control over the system. Over the centuries, we have tried a lot of ways to regulate finance, even executing failed bankers \citep{Kohn99}. Still, that did not work.

A financial authority entrusted with overseeing the financial system has two objectives. The first is microprudential regulation, or micropru, which is concerned with day-to-day issues such as risk management, consumer protection, and fraud. AI generally helps micropru as data is ample, the rules are mostly fixed on the timescale decisions are made, and the cost of mistakes is small. Most financial regulations, including substantial parts of the Basel Capital Accords and securities regulations, fall under micropru. 

The second policy objective is macroprudential regulations, or macropru, focussed on broad-picture issues relating to financial stability. Macropru looks at the long run, reducing the likelihood of large losses, stress and crises, and when such events happen, resolving them in the most expedient way possible. Macropru is much more difficult to execute and less accurate than micropru, as events to be controlled are infrequent and often unique. Furthermore, macropru is generally much more political, with the most serious events which overseen by the political leadership. While the scarcity of data coupled with the mutability of rules and unclear responsibility always frustrates the macropru authorities, it poses particular challenges to AI due to its particular dependence on data. That implies considerable uncertainty about the ability of AI to advise on and implement macropru and gives rise to new dangers, as noted by Danielsson, Macrae and Uthemann (2022)\nocite{DanielssonMacraeUthemann2019}.

Below, we identify five channels of financial vulnerabilities that any authority, whether working for the government, outside standard bodies, or internal control, has to contend with when attempting to exercise control over activities in the financial system. While each of these has always been present, they pose particular challenges in a financial system where AI is used extensively both by the public and private sectors. 

\subsection{Usefulness of data}\label{sec:data}

Data should play to AI's advantage as the financial system generates many petabytes of it daily. Every transaction is noted, all decisions documented, decision-makers are monitored and recorded, and we can track processes over their lifetime. Financial institutions must report some of this data to the financial authorities, and the authorities can demand almost all of it later. One might expect this ocean of data would make it easy for AI to study the financial system in detail and identify all the causal relationships. That is mostly true micropru but not macropru.

Start with the basic measurement. The standards for recording financial data are inconsistent, so different stakeholders might not observe particular activities in the same way, leading to complex matching problems. Identification coding and database design can differ significantly. Financial institutions have a lot of legacy systems not set up with data collection and sharing in mind, rendering data collection, especially in a format that is standard across the industry and necessary for the authorities, costly and error-prone. Fortunately, while real today, these problems are rapidly being overcome, not the least with the aid of AI. 

A bigger challenge is all the silos in the regulatory structure that hinder data sharing. Most data stays within a financial institution and is not shared. Even when shared, the financial institution might retain copyright, allowing it to control access to it so that data might be available for compliance but not broader objectives such as financial stability. Furthermore, financial system data are collected by authority silos where data sharing is limited. There might be restrictions on data sharing between the supervisory and statistical unit of a central bank, between authorities in the same country or between jurisdictions. These problems were made clear during the 2008 crisis when nobody had an overview of the aggregate market for structured credit. The situation has improved somewhat since then, for example, thanks to mandatory trade reporting. However, silos persist and are likely to continue hindering data sharing.

Finally, when it comes to the most serious events, systemic financial crises, the events under consideration are, fortunately, rare. The typical OECD country only suffers a systemic crisis one year out of 43, according to the crises database maintained by \citet{LaevenValencia2018}. However, that fortunate low frequency frustrates the data-driven analytics AI depends on. 

These three issues, data quality, silos and rarity of events, are usually not all that important for microprudential regulations. This is not so for macropru, where they amplify each other, negatively impacting the design of macroprudential regulations, enforcement and crisis resolution.

\subsection{Unknown-unknowns}

The same three fundamental vulnerabilities cause every financial crisis. Financial institutions use high degrees of leverage, rendering them vulnerable to shocks, self-preservation in times of stress, which makes market participants prefer the most liquid assets, and system opacity, complexity and asymmetric information, causing market participants to mistrust each other in times of heightened uncertainty. However, these three vulnerabilities are high-level and abstract, and every crisis is unique in detail. It is almost axiomatic that the macropru authorities are surprised by crises. It could not be any other way because the authority otherwise would have taken precautionary actions that would have avoided the crisis in the first place. This means that the most severe financial crises are by definition unknown-unknowns or uncertain in Frank Knight's (1921) \nocite{Knight21} classification.

The uniqueness of crises creates particular problems for the designers of macroprudential regulations because they generally only learn what data is most useful after a stress event. That was the case in 2008, where we only realised afterwards that sub-prime mortgages being put into structured credit products --- the banks held onto the most senior and junior tranches where risk modelling was poor --- was the key channel for the crisis. Obvious after the event but practically impossible to discover before. When the analyst faces an almost infinite number of signals and an enormous number of false positives, it is very difficult, to the point of impossible, to identify which data is useful until after a crisis event is already underway. It is too late to have preventative regulations in place by this time.

While the authorities can scan the system for specific causes of vulnerabilities, their job is frustrated by the almost infinite complexity of the system. The supervisors can only patrol a small part of that infinitely complex space. Even if the supervisors, AI or human, could monitor all threat vectors and assign a probability to each --- an impossible task --- they still have the problem of picking notification thresholds and identifying the type 1 and type 2 errors of the model. Doing that comprehensively is an impossible task, demanding almost infinite resources. The system's complexity and measurement noise mean that the number of notifications would be very large, with mostly false positives. Furthermore, such intrusive monitoring might sharply curtail desirable risk-taking because of the false positives and be seen as socially unacceptable. 

AI both helps in identifying the unknown-unknowns and drives their creation. The impact of AI, therefore, depends on which of these two has a stronger impact. We fear the latter factor will dominate.

\subsection{System responses}\label{sec:sys}

A key challenge for AI working on macropru, but not generally for the microprudential authorities, arises from the dynamic interaction of the financial system participants. A helpful framework for understanding the problem is the Lucas (1976) critique\nocite{Lucas76}, which states that the decision rules used by economic agents depend on the underlying economic environment and can change as regulations change, undermining their effectiveness. Goodhart, in his critique, applies this to regulations, stating, ``Any observed statistical regularity will tend to collapse once pressure is placed upon it for control purposes.'' Changes to financial regulations or the level of supervision, including changes to the crisis resolution playbook, as well as decisions taken during resolution, will change the responses of the private sector to similar crises in the future in a way that is unexpected.

A particular consequence is the measurement of financial risk, an essential task for any regulator. Besides the data issues discussed above, there are particular technical issues for why risk measurements can be misleading. It is helpful to use the classification scheme proposed by \citet{DanielssonShin2002}, which separates financial risk into two categories: exogenous and endogenous. Exogenous risk emphasises risk measured by statistical techniques based on historical outcomes in financial markets, typically prices. Endogenous risk, by contrast, captures risk that arises from the strategic interaction of the economic agents that make up the financial system. 

Exogenous risk is easy to measure, and AI excels at it. Identifying endogenous risk is difficult because it captures outcomes only visible in extreme stress when self-preservation and mistrust of counterparties are crisis amplifiers. AI will help identify and cope with some forms of endogenous risk. In times of stress, humans might specify the objectives, while AI might be better at reacting to a crisis given those objectives. Even then, it will have to contend with the socially damaging individual optimal mistrust and self-preservation. Ultimately, an AI engine interacting with another AI engine would likely face similar issues as with humans facing humans today, pre-emptive runs, multiple equilibria and the like. These outcomes happen even in fully rational settings, and AI-advised and managed systems might be more susceptible to such outcomes than more human-centred arrangements.

The relative importance of exogenous versus endogenous risk depends on the problem. For most microprudential regulations, the frequency of events and lack of large strategic interactions mean that assuming risk is exogenous is usually quite innocuous. However, any authority using data-driven analytics that uses exogenous risk measurement to assess the risk of financial instability will likely be seriously misled as to how the financial system evolves in times of stress, as noted by \citet{DanielssonJamesValenzuelaZer2015a}.

\subsection{Pre-specified objectives and distributed decision making}
\label{sec:objectives}

While both humans and AI can operate in an environment with mutable objectives, both become less effective and more prone to mistakes as events become infrequent. The worst case is when the objective is unknown ex-ante and cannot be learned, when its potential for providing seriously flawed advice and catastrophically bad decisions is the strongest. 

Mutable objectives do not pose much of a problem to micropru. The rulebook is known, and usually static on the timescale decisions are made. While it evolves in response to events and the response of the regulated to regulations, AI can quickly update its understanding of the objectives by adopting changes to the rulebook, by observing how the human supervisors act.

This is different in macropru, as it operates on very long timescales. The time between relevant events is usually in decades. It can be very difficult to define the macroprudential objective except at the highest levels of abstraction, such as the prevention of severe dysfunction in key financial markets and especially the failure of systemically important institutions. This applies to the design of regulations, supervision and crisis resolution. It is not easy to make a case for the need to allocate significant resources to prevent something that only happens in the distant future. Meanwhile, macroprudential regulations are susceptible to lobbying and political interference, which means that the objectives of regulations change over time. While it is straightforward to set the policy objective to ``prevent the failure of SIFIs'', the actual implementation of such a policy has widespread implications on the entire financial system, and the authorities making such decisions in real-time will draw on facts and interests only visible at the time. Learning from previous crisis interventions might be of little help as the concrete goals the actions under consideration are to achieve are likely specific to particular circumstances and political environments.

The most severe financial crises can have catastrophic consequences if not addressed adequately, with direct economic costs in the several trillions of dollars, as noted by \citet{BarnichonMatthesZiegenbein2022}, with a large number of people materially affected, leading to considerable political consequences as well. When such crises happen, society demands we do what it takes to resolve the crisis. \citet{ChwierothWalter2019} note that bailouts have become what they call a middle-class good, making them inevitable. 

Often, the rules and the laws in place stand in the way of the most effective crisis resolution. Emergency sessions of Parliament to rectify that are not uncommon, such as Switzerland's resolution of Credit Suisse.\footnote{https://www.bloomberg.com/news/articles/2023-03-20/credit-suisse-collapse-reveals-some-ugly-truths-about-switzerland-for-investors} Emergency constitution clauses might be invoked. \citet{Pistor2013}, in her legal study of the resolution of financial crises, finds that if the existing law prevents the most effective course of action, there is acceptance from the political and judicial system to suspend the law in the name of the higher objective of crisis resolution. Furthermore, when a severe crisis happens, the political leadership takes charge, inevitably because if it becomes necessary to change or bypass the law or significantly redistribute resources, the political leadership is the only entity with the necessary legitimacy.

Given the fluidity of this process, will AI find it particularly difficult to provide real-time advice? We have a long experience in resolving crises and a relatively good understanding of the process. The regulatory system is usually modular, with separate authorities and fiercely guarded mandates. In the most severe crises, these silos break down. All relevant authorities, the affected private sector, the judiciary, and especially the political leadership have come together to decide how to resolve the crisis. Government ministers usually lead this process. This may involve the same entities in other financial centres in a global crisis. Each stakeholder brings their own education, ethics, professional experience and objectives to the table. Such a process can be highly robust. All pertinent issues are discussed, including information that was confidential or implicit until then. Such analysis depends on implicit knowledge and intuitive understanding its participants have of the current situation and each others' views and objectives. \label{page:expert}Intuition can mean finding analogous problems from seemingly unrelated domains that can provide creative solutions for the current crisis that has not been encountered before. This process of successful extrapolation is often vital to crisis resolution and can imply ignoring or revaluing previously pursued goals. There are many such examples in history.\label{page:expert} The German central banker Hjalmar Schacht used short squeezing to stop the hyperinflation in 1923. The Swedish central bank in 1992 created the good bank-bad bank model for crisis resolution. Suppose AI is to become an effective macropru regulator. In that case, it will have to reproduce such a process and pay attention to nuance, hidden objectives and mutable objectives.

Crisis resolution is arguably the most important aspect of financial policy, especially for central banks --- one of their raisons d'\^etre. Nevertheless, given the above challenges --- scarce data, unknown-unknowns, endogenous structural changes in response to attempted control --- humans struggle with this task, and this is where AI could benefit them the most. Unfortunately, it has to overcome the same conceptual problems, and this challenge is as difficult for AI as for human regulators, if not more so. 

Paradoxically, progress in AI's suitability for the task might come from a better human understanding of these problems that can then be translated into better algorithms.

\subsection{Incentives}
\label{sec:Incentives}

The objectives of the various market participants are specific to them and often in conflict with those of other participants and society. In the past, we have seen cases where market participants actively and deliberately create heightened stress since anybody forewarned of stress can profit. Creating and amplifying crises is profitable. \citet{Lowenstein00} shows just one example from the LTCM crisis in 1998, and what is more frequent, their individually optimal actions --- run for safety, survival --- endogenously amplifies socially undesirable stress. That is why we have financial regulations: to align the private sector's incentives with society. The interactions between financial supervisors and the private sector decision-makers can be seen as a principal-agent (PA) relationship, where the highly resourced and sophisticated private sector --- the agents --- wants to get on with successfully running their businesses. In contrast, the supervisors --- the principals --- aim to ensure that the agents operate a financial system that efficiently channels savings to investments while not abusing their clients too much or taking excessive risks that can cause costly crises. 

The PA problem applied to financial markets is particularly difficult for three reasons. First, contracting is incomplete, meaning we cannot specify what will happen in every possible contingency. We might not even know what the contingencies are. Especially as the very behaviour to be controlled, risk-taking, means there are circumstances that the principal does not anticipate. Risk is a latent variable that cannot be measured directly but only imprecisely inferred from market movements. Consequently, it can be difficult to determine whether an undesirable outcome is due to misbehaviour or innocent bad luck. Second, because implicit or explicit government guarantees are fundamental to all financial regulations, market participants, knowing that some of their risk-taking is publicly insured, can behave in a way that makes government bailouts more likely, creating moral hazard.\footnote{\citet{dewatripont1994prudential} provides a detailed analysis of the problem of financial regulation with incomplete contracts, government guarantees and private sector moral hazard.}

As we increasingly use AI in regulations and the private sector, a new dimension in the PA problem emerges. The one-sided financial institution-regulator problem becomes two-sided, with the addition of the relationship between the regulators and financial institutions as principals and their respective AI as agents. The reason we worry about the AI's hidden actions is that objectives are misaligned. This means the human operator of AI needs to incentivise its AI to act in humans' interest, so there is also a principal-agent problem between the agent and its AI.

The presence of the two-sided PA problem amplifies the already significant existing PA problems in financial regulation. The authorities will have to explicitly consider how to align the AI engines, whether directly operated by the regulated or by their AI vendors, with the objectives of the authority. Existing mechanisms for managing the regulator-bank PA relationship will not be adequate when AI is involved. In the technical terminology of moral hazard, AI has more hidden actions at its disposal and will command higher information rents. While AI's objectives might be simpler, its actions/strategies to achieve the objectives may be more complex. It leads to difficult questions. What higher-level objectives does this loss function induce, and how do we formulate an incentive scheme that imposes the correct constraints on AI behaviour? How can we make it internalise negative externalities, such as the possibility of deliberately creating fire sales or runs to trigger privately beneficial but publicly costly bailouts? \section{Artificial intelligence}\label{sec:AI}

There is no standard definition of artificial intelligence (AI). One of the pioneers of AI, John McCarthy, saw it as ``the science and engineering of making intelligent machines''\footnote{jmc.stanford.edu/artificial-intelligence/what-is-ai/index.html} where perhaps ``intelligence measures an agent's ability to achieve goals in a wide range of environments'' in the definition of \citet{legg2007universal}. ChatGPT defines it as ``The capability of a machine to imitate intelligent human behaviour, including learning, reasoning, problem-solving, perception, and language understanding''. We see it as a computer algorithm performing tasks a human would otherwise do, \citep{RussellNorvig2010}. In Russell's (2019) \nocite{Russel2019} taxonomy, AI as a rational maximising agent resonates with the economic notion of utility maximising agents and hence is particularly helpful in the analysis of AI use in the financial system.

While statistical and machine learning\footnote{ML provides the fundamental building blocks of most AI, finding statistical regularities in large data sets and helping provide AI engines with a deeper understanding of the underlying latent structure that generates the data of a given problem domain. See the Appendix for details on ML that are useful in financial applications. } (ML) algorithms are limited to giving quantitative inputs for humans making decisions; AI also provides recommendations and even makes decisions. When we charge AI with a task, it searches for the best course of action given its objectives and understanding of the problem domain. A key differentiator between the conventional and AI way of operating is how AI learns in real-time while interacting with the environment.

\subsection{From reactive machines to AGI}\label{sec:AI:AGI}

Even though the ability of AI engines has increased rapidly, today's AI merely matches patterns and does not understand them in the same way as humans. It is unclear what achieving such an understanding will involve. When considering the abilities of various AIs, including those only hypothesised, it is beneficial to classify AI based on its abilities. That is a surprisingly difficult task since the experts do not agree on classification, terminology, ability or future potential. At the risk of oversimplification, there are four types of AI. 

The first is often termed a \textit{reactive machine}, which is a system with no memory of previous decisions and designed to perform a specific task. In other words, it only works with already available data. Typical examples are classification engines that identify objects and photographs and recommendation engines.

The second category is \textit{limited memory} AI, which knows what happened in the past, keeps track of past actions and observations and can condition its response to a given situation on these past experiences. Generative AI, such as ChatGPT, falls into this category, as it is a system that predicts the next word or phrase or some other object based on the content it is working with. Other examples include virtual assistants and self-driving cars. Time series models are also of this type. 

We then have a \textit{theory of mind}, not yet realised, AI that aims to understand thoughts and emotions and consequently simulate human relationships. They can reason and personalise interactions. Importantly, from the point of view of financial policy, a theory of mind AI should be able to understand and provide context for policy decisions, which today's AI cannot do, and hence be invaluable in crisis resolution. AI that understands how its interlocutors think can explain its decisions to them and reason about how they will react to its actions and plan accordingly. 

And finally, \textit{artificial general intelligence} or AGI, which is still purely theoretical while often hypothesised. It would use existing knowledge to learn new things without needing human beings for training and being able to perform any intellectual task humans can. If it reaches that level, increased computer capacity will allow it to surpass humans.

The potential for the eventual emergence of a theory of mind and AGI AI remains controversial, and experts have strong disagreements on both the likelihood of such eventualities and what it takes to reach them. The ability of an AI engine depends on three factors: network structure, data and compute. The more relevant data we feed into the engine and the more compute we apply, the better it will perform. What is controversial is whether that is all we need to make progress on the theory of mind and AGI or if a conceptual leap in the ways of learning is also required. 

A further point of controversy is whether the next token prediction is sufficient for achieving AGI, since in that case, all that is required for AGI is a sufficient amount of training data and compute, conceptual improvements in network technology are not needed.

\subsection{How do we know AI does what it is supposed to?}\label{sec:AIusefulness}

One of the hardest problems for AI applied to advising and making decisions in complex social settings, like financial policy, is the specification of its objectives. Why should we trust its analysis and explanations, and how to make AI systems behave in line with human intentions and values?\footnote{For a survey of AI alignment work, see, e.g., \citet{ji2023ai}, and \citet{shevlane2023model}, and \citet{bengio2023managing} model evaluation for extreme risks.}

A key challenge is that the training algorithms need to know what to optimise for since AI need well-specified objective functions that evaluate the cost and benefits of alternative courses of action given the current state of the environment and its future evolution. That extends to AI taking into account how the system reacts to its actions, which is particularly important in financial applications. Misspecification, and especially insufficient specification of problems, leads to suboptimal and even catastrophically bad decisions. 

In the 1980s, an AI decision support engine called EURISKO\footnote{Following the passing of EURISKO's creator, Douglas Lenat, the source code was re-discovered and put online, blog.funcall.org/lisp/2024/03/22/eurisko-lives} used a cute trick to defeat all its human competitors in a naval wargame, sinking its slowest ships to maintain manoeuvrability. This early example of AI reward hacking, something humans are experts in, illustrates how difficult it is to trust AI. How do we know it will do the right thing? Human admirals don't have to be told not to sink their own ships, and if they do, they either have high-level political acquiescence or are stopped by their junior officers. Any current AI making autonomous decisions has to be told, or learn from observing human decisions, that sinking its own ships is not allowed. The problem is that the real world is far too complex for us to train AI on every eventuality. AI will predictably run into cases where it will make critical decisions in a way that no human would. EURISKO's creator, Douglas Lenat, notes that ``[w]hat EURISKO found were not fundamental rules for fleet and ship design; rather, it uncovered anomalies, fortuitous interactions among rules, unrealistic loopholes that hadn't been foreseen'' \citep[][p 82]{Lenat1983}. Each of EURISKO's three successive victories resulted in rule changes intended to prevent repetition. Still, in the end, the only thing that worked was telling Lenat that his and his AIs' presence was not welcome. 

If we ask ChatGPT whether it is okay for admirals to sink their own ships, it gives a well-grounded and nuanced answer. However, this is a well known example, likely in the canon ChatGPT trained on. Would it have come up with the same answer if Lenat had not entered EURISKO in the naval board game? We don't know.

An AI engine can lead to suboptimal outcomes that are harder to detect than in the EURISKO case. Having control of some system and being given the objective of forecasting, it might attempt to manipulate the system's structure to meet the objectives as it sees them. This can have severe consequences for microprudential and macroprudential regulations, as discussed in Section \ref{sec:channels} below. 

That means it is essential that AI can explain its reasoning, just like humans are required to. That can be difficult since AI outputs are simply a nonlinear mapping of input data, where decisions are affected both by the network structure and trained parameters. The machine may say yes or no based on one particular configuration of a trillion parameters. 

AI, which cannot explain its reasoning, can only be used for some applications. Not surprisingly, considerable resources are being brought to bear on that problem, allowing the operators to trace particular recommendations to the input data. Instead of a recommendation being merely a function of model parameters, recommendations are determined by particular inputs that are analysed logically. Benchmarking AI engines to specific tasks could be particularly valuable in financial applications, as we discuss below.

\section{The channels for how AI can destabilise the financial system}\label{sec:channels}

The reason why AI increases systemic risk is, perhaps paradoxically, its very strength and efficiency. In the words of \citet{bengio2023managing}, ``Compared to humans, AI systems can act faster, absorb more knowledge, and communicate at a far higher bandwidth. Additionally, they can be scaled to use immense computational resources and can be replicated by the millions.'' The financial system has a large number of potentially destabilising feedback mechanisms that are created by the interaction of actors with significant financial resources that are maximising profits in regular time and the probability of survival during stress. The combination of AI that excels at discovering optimal strategies, the system's complexity, highly resourced strategic actors and the inherent destabilising factors already prevalent in the financial system viciously reinforce each other.

In our analysis of how AI can destabilise the financial system, we build on extant work on AI safety, like \citet{weidinger2022taxonomy}, \citet{bengio2023managing} and \citet{shevlane2023model}, which identifies several societal risks arising from AI use, including malicious use, misinformation and loss of human control. In addition, AI brings a particular market structure. When adapting these AI risks to the financial sector and analysing them in the context of the financial vulnerabilities developed above in Section \ref{sec:conceptual}, we arrive at four channels for how AI can destabilise the financial system.

\subsection{Malicious use of AI}

The first channel arises because the system is replete with highly resourced profit-maximising economic agents not too concerned about the social consequences of their activities. They seek to bypass controls and change the system in ways that benefit them while being difficult for other market participants and regulators to detect. Technology has always helped such agents, and they are already embracing AI. Financial institutions and their staff deploy considerable resources in their pursuit of profit and usually are not very concerned about the wider social impact of their activities. They can change the system in ways that benefit them while not being detectable by others, as we have seen many times in the past. AI provides such actors with new opportunities, perhaps via adversarial attacks by feeding particular data into others' training algorithms or influencing market structure. That might not even be necessary since it is straightforward to exploit for private, even illicit, gain gaps in how AI sees its responsibilities where ``maximise profit'' leads to market manipulation without explicit instruction by humans.

One way such malicious use might occur is by manipulating governance processes. The financial authorities and internal control aim to align the interests of system participants with those of the organisation that employs them and society. As AI is particularly effective in finding profitable loopholes and amplifying vulnerabilities, it can facilitate misbehaviour that, while legal, is damaging both to society and even the institution employing the AI.

The operator of the AI could take that one step further and use it to deliberately create market stress, which, of course, is profitable to those forewarned. We have seen many non-AI examples of such conduct in the past, and a key purpose of securities and macroprudential regulations is to prevent such behaviour.

We expect the most common malicious use of AI will be by employees of financial institutions, who are careful to stay on the right side of the law. AI will also facilitate illegal activities, such as rogue traders and criminals, as well as terrorists and nation-states aiming to create social disorder. The financial system has always attracted such hostile agents, and nation-states view financial system attacks as one part of their arsenal, as we have witnessed many times in the past. The inherent vulnerability of the financial system to adversarial attacks and the ease of manipulating AI engines create potential for particular attack vectors that can be impossible to identify ex-ante. One example comes from \citet{hubinger2024sleeper}, who identify what they call a ``sleeper agent'' that usually acts as expected but gives deceptive answers when given special instructions. Here, AI engine corruption becomes another form of cyber attack. With such an attack vector in place, an AI engine becomes a tool for well-resourced hostile agents that might be difficult or impossible to prevent in real-time, facilitating nation-state attacks on critical infrastructure. That is particularly relevant when a superior ability to identify optimal timing strategies allows them to solve the problem of double coincidence, as discussed in \citet{DanielssonFoucheMacrae2016o}, creating heightened system fragility, perhaps by manufacturing a liquidity crisis as an attack amplifier.

\subsection{Misinformed use and overreliance on AI}\label{sec:misinformed:over}

The second channel emerges when the human operators of AI engines don't fully appreciate how AI operates and then end up using it inappropriately. As we come to trust AI analysis and decisions and see how well it performs in increasingly complex and essential tasks, it builds up trust. There are many examples where technology was initially met with scepticism but then became trusted, with just some recent examples: computers landing aeroplanes, doing surgery and driving cars. While advice from AI might be reviewed by humans, once we see it suggest the right thing many times, eventually, it becomes accepted. When AI performs at a level equal to or better than its human counterparts but at a much lower cost, it will be welcomed by senior management. Its very success creates trust. As that trust is earned on relatively simple and safe repetitive tasks and seeing AI succeed in more complex jobs, it will be asked to take on ever more sophisticated tasks.

That is particularly relevant when AI is used in domains with unclear objectives, which is very common in the financial system. As events become less frequent and more serious, data-driven decisions increasingly involve extrapolation, interpretation, intuition and nuance, none of which AI in its current form is good at. As we discuss on Page \pageref{page:expert}, many crises have been resolved by very creative solutions that had never been used before.

Because successful AI architectures are data-driven engines, their analysis is predominantly founded on statistical regularities. Such AI engines are designed to provide advice and responses to prompts, even if they have very low confidence about the accuracy of the answer. Of course, many architectures are probabilistic and can provide measures of statistical accuracy. They can even make up facts or present arguments that sound plausible but would be considered flawed by an expert, both instances of the broader phenomenon of AI hallucination. The risk is that the AI engines will present confident recommendations about outcomes they know little about. 

\subsection{AI misalignment and control avoidance}

The third channel for how AI can be destabilising relates to difficulties in aligning the behaviour of AI engines with the objectives of both their owners and financial authorities. The latter is particularly challenging if the objectives of financial regulations cannot be defined precisely, as usually is the case, as we discussed in Section \ref{sec:objectives}. Problems of misalignment arise when AI makes decisions because of the impossibility of fully specifying all objectives it has to meet, a form of incomplete contracting. The best solution an engine comes up with might be undesirable for the institution it works for. While such individual misbehaviour is certainly all too common in today's human-centred financial institutions, it becomes more prevalent as AI use increases. Humans have been taught ethics and have general notions such as ``don't cheat, steal, no violence.'' Most organisations try to recruit individuals for whom that education was successful. AI might require a similar broad education if its objectives cannot be fully specified ex-ante.

\citet{scheurer2023technical} provides an example of how individual AI can spontaneously choose to break regulations in their pursuit of profit. Using GPT-4 to analyse stock trading, they told their AI engine that insider trading was unacceptable. When they then gave the engine an illegal stock tip, it proceeded to trade on it and lie to the human overseers. Here, AI is simply engaging in the same type of illegal behaviour so many humans have done before.

The problem of misalignment is particularly difficult when several AI engines interact. For example, market manipulation is most successful when several investors collude. Similarly, the payoff of many trading activities, such as carry trades and short sales, increases with the number of investors participating in them. A particularly damaging consequence of the strategic complementarities prevalent in the financial system relates to the objectives of financial institutions. They maximise profit most of the time, perhaps 999 days out of a thousand, but optimise for survival during stress. The behaviour on that one day out of a thousand is particularly damaging when financial institutions seek safety by withdrawing liquidity, which can lead to destabilising dynamics such as fire sales, bank runs, and credit crunches. The very high level of AI performance can, perhaps paradoxically, increase the likelihood of damaging coordinated behaviour. They might find better strategies for exploiting strategic complementarities for private gain at the risk of lowering market quality or even threatening financial stability.

An example of AI collusion is \citet{CalvanoEtAl2020}, who find that independent reinforcement learning algorithms instructed to maximise profits quickly converge on collusive pricing strategies that sustain anti-competitive outcomes. It is much easier for AI to behave in this collusive way than humans, as such behaviour is very complex and often illegal. AI is much better at handling complexity and is only aware of the legal nuances if explicitly taught or instructed.

The most difficult problems of misalignment arise due to the interaction of the various AI working throughout the industry. Long before computers got involved with the financial system, humans coordinated damaging behaviour due to the misalignment of individual short-term incentives with the long-term incentives of the same market participants and society at large. The crises of 1763, 1914 and 1929 \citep[see, e.g.][]{DanielssonBook2020} are just two of many such examples, as is the tulip mania in the Netherlands in the 1630s, see \citet{AliberKindleberger2015}. 

The use of computers amplifies the potential for such undesirable outcomes, as two recent examples illustrate. The largest stock market crash in history happened in 1987 when global markets fell by 23\% in one day. The reason was trading algorithms that, in a way not foreseen, coordinated in creating vicious feedback between prices collapsing and algorithm-induced selling, as documented by \citet{gennotteleland90}. Similarly, the quant crisis in the summer of 2007, as noted by \citet{KhandaniLo2007}, was due to trading algorithms trained to sell on falling prices but not being told of the potential for aggregate feedback caused by trading algorithms in a number of funds coordinating on selling. Such factors have further been at the root of the various flash crashes. AI will likely amplify such problems of misalignment. AI might implement cross-market strategies that no humans will be able to understand, let alone unwind. While the traditional approach to algo (or human) induced market crashes is trading stops, i.e., a simple kill switch, that might not be feasible for complex cross-market AI strategies.

\subsection{Market structure}\label{sec:oligopoly}

Monoculture is an important driver of booms and busts in the financial system. As financial institutions come to see and react to the world in increasingly similar ways, the more they coordinate in buying and selling, leading to bubbles and crashes or, even worse, credit booms and credit crunches. aucomm{For me, credit booms and crunches can be instances of bubbles and crashes.} Banks' capital calculation is a particular example, not the least via risk weights. Such procyclical behaviour has always been a feature of financial markets, long before the use of computer technology. Still, it is amplified by technology. 

AI will likely exacerbate this already strong channel for financial instability. It excels in finding the best methods for measuring risk by using generative models that combine a scientific understanding of the nature of risk with stochastic processes of financial assets. While that might lead to the best possible measurement of risk, it also has the unfortunate consequence of driving market participants to see risk in the same way --- beliefs. AI then will find the best practices for managing risk --- action. The consequent harmonisation of beliefs and action is strongly procyclical, as financial institutions react to shocks similarly, excessively amplifying buying and lending, inflating bubbles, and eventually selling rapidly and withdrawing credit.

The oligopolistic nature of the AI analytics business further strengthens that efficiency-induced procyclicality. ML design, input data and compute affect AI engines' ability. These are controlled mainly by a few technology and information companies, which continue to merge, creating an increasingly oligopolistic market. The median salary for specialists in data, analytics and artificial intelligence in US banks was \$901,000 in 2022 and \$676,000 in Europe.\footnote{https://www.bloomberg.com/news/articles/2023-11-28/goldman-raided-by-recruiters-in-wall-street-fight-for-ai-talent} Since there is a considerable shortage of the necessary human capital and the productivity of those experts is directly affected by the network effects of working with other experts, as well as the data and compute available to them, this alone puts designing the most effective AI engines out of the reach of all but the largest financial institutions. Furthermore, the currently largest neural networks have trillions of parameters, GPT 4 with 1.7 trillion, requiring significant funding and access to specialised data centres with the requisite GPUs. The current top engines have hundreds of millions of dollars in annual budgets. OpenAI reported that the cost of training GPT-4 exceeded \$100 million.\footnote{https://www.wired.com/story/openai-ceo-sam-altman-the-age-of-giant-ai-models-is-already-over}

Finally, financial data vendors have concentrated considerably over the past few years, with only a few large vendors left, such as S\&P Global, Bloomberg and LSEG. They have unified databases with standardised labelling and graph structures linking disparate categories of data, including proprietary and not publicly available data, facilitating engine training. It is a concern that neither the competition nor the financial authorities have fully appreciated the potential for increased systemic risk due to oligopolistic AI technology in the recent wave of data vendor mergers.

Taken together, AI-driven financial analytics is characterised by increasing returns to scale technologies with high entry costs, making an oligopolistic market structure highly likely. In such an environment, outsourcing analytics to one of those vendors might be the best, maybe even the only feasible solution for all but the largest financial companies. Anecdotal private sector evidence indicates that many financial institutions already prefer to outsource data-driven applications to specialists. A particular example is risk management as a service, RMaaS, with a leading vendor being BlackRock's AI-fuelled Aladdin. 

The risk is harmonised beliefs and actions, driving procyclical behaviour and exposing users to the same blind spots.

\section{The use and regulation of AI}\label{sec:GoodUse}\label{sec:regulation} 

The financial system has always been an early and enthusiastic adopter of technology. One of the first applications of the first transatlantic telegram cable in 1858 was the transmission of stock prices. Nathan Rothschild supposedly used pigeons to get the first news of Napoleon's defeat at Waterloo in 1815 in order to manipulate the London stock market. As many such examples show, technology is both a source of efficiency and risk, and we now see the same with AI.

Most public and private financial system applications can be supported by the two forms of AI accessible today, reactive and limited memory, and their associated architectures. As the engines' capabilities improve, and we come closer to a theory of mind AI, and even AGI, we expect AI use to expand in both the public and private sectors \citep{turing2023,KinywamaghanaSteffen2021, Korinek2024}.

Most job functions in the public and private parts of the financial system are quite routine, and there is no technical reason why AI today cannot perform the majority of these jobs. Credit allocation, customer interactions, report generation, pricing of insurance contracts, fraud detection, compliance and risk management are just a few of the tasks where existing AI have proven their worth. While many financial institutions are making large investments in AI, what often holds AI adoption back is lack of familiarity with it and insufficient access to human resources, data and compute, and perhaps most importantly, caution based on worries about what sort of mistakes AI advice and decisions could lead to. Many financial institutions lack the necessary resources, and smaller legacy institutions can find AI particularly challenging. Legacy systems and processes might not be well suited to AI adoption.

These issues don't generally apply nearly as much to challenger financial institutions, such as the various neo-banks. These are new institutions with modern technology stacks, benefitting from technology staff and senior management that are conversant with modern technology. In a competitive market, greater efficiency and better delivery of financial services due to AI will push its adoption across the private sector.

Since the financial system is one of the most heavily regulated parts of the economy, the financial authorities similarly take a keen interest in how technology is used. They want to identify the threats technology poses, how to harness it to improve the system and how to regulate it. A particular challenge for the architects of financial regulations is that they simultaneously need to keep the system safe and make it more efficient. As the private sector increasingly adopts AI, the supervisors will have to address both how it changes the behaviour of private sector institutions and how to regulate AI engines in use.

Regulating the financial system is not an easy task. The authorities have to balance conflicting objectives while wielding limited resources when patrolling a nearly infinitely complex financial system. Once the authorities are focused on particular activities, the resulting risk might be contained, but what tends to happen is that the forces of instability emerge elsewhere. Risk is like a balloon; squeeze it in one area, and it expands elsewhere. Indeed, it is almost axiomatic that instability arises where the authorities are not looking. And there is plenty of scope for instability in an infinitely complex system. The authorities will have to contend with how AI changes regulations, how to regulate AI and perhaps most importantly, how to harness it in their operations.

\subsection{The authorities' use of AI}\label{sec:use:auth}

Most financial authorities are approaching AI cautiously and are behind the private sector in both analysing how AI changes the system and how to use it in their activities. Anecdotal evidence suggests that their current AI use is mostly limited to routine forecasting, basic analysis with LLMs and low-level data processing. Some authorities are increasingly considering how AI will affect their mission, as evidenced by \citet{LeitnerSinghKraaijZsamboki2024}, \citet{Moufakkir2023} and \citet{Cook2023}. If they don't keep up, the authorities risk becoming increasingly irrelevant if their way of working, such when monitoring, regulating and supervising the financial system, is not geared towards private sector use of AI. The financial authorities find AI challenging. Engaging with it necessitates investments in data, human capital and compute, as well as demanding re-thinking in how to regulate finance. 

The most straightforward is data, as authorities already have access to large databases, models and technical processes, including confidential disclosures, that can be used to train AI. This includes:
\begin{enumerate}
 \item Observations on past compliance and supervisory decisions
 \item Prices, trading volumes and securities holdings in fixed-income, repo, derivatives and equity markets
 \item Assets and liabilities of commercial banks
 \item Central clearing activities
 \item Network connections, like cross-institution exposures, including cross-border
 \item Textual data 
 \begin{enumerate}
 \item The rulebook
 \item Speeches, policy decisions, staff analysis
 \item Records of past crises resolution
 \end{enumerate}
 \item Internal economic models 
 \begin{enumerate}
 \item Interest rate term structure models 
 \item Market liquidity models 
 \item Inflation, GDP and labour market forecasting models
 \item Equilibrium macro models
\item System stability models
 \end{enumerate}
\end{enumerate}

By contrast, the authorities can find it difficult to acquire the necessary human resources and compute, not the least because of the very high costs we discuss above in Section \ref{sec:oligopoly}. However, because many of the largest costs are associated with training general-purpose engines, it is easy to overstate those problems. The marginal cost to the authorities is likely to be significantly smaller. To begin with, they may be able to extend extant engines via transfer learning on curated datasets for financial sector use that include the items listed above. Furthermore, open-source platforms running on relatively low-end computers are sufficient for many applications.

There are many activities within the financial authorities where AI would be of considerable help. Start with the design of microprudential regulations. ML can be used to translate the rulebook into a logic engine, allowing analysis of internal consistency and coverage and facilitating AI supervision, and then enforcing compliance with it. Several authorities have already made progress in this area. Undesirable behaviour, including that not covered by extant rules, can be simulated in order to make the rulebook robust against such outcomes. AI can help in supervision, providing routine compliance advice to the private sector, recommending regulatory actions, and even making supervisory decisions. While human supervisors would initially closely oversee it, reinforcement learning with human feedback will help supervisory AI to become increasingly performant, and repeated success will build trust. This could be useful in resolution, such as finding the least-cost resolution by evaluating a number of alternative resolution approaches. For challenging areas of microprudential supervision, some types of AI might be particularly beneficial in understanding complex areas of authority-private sector interactions, such as fraud detection and money laundering. Hostile agents make extensive use of AI to hide their activities and find new areas to exploit. The authorities, along with private sector advisory firms, are actively counteracting that with AI. 

AI will also be useful in other routine activities, such as forecasting and ordinary economic and system fragility analysis. That may use public and private data, established economic theory embodied in economic models and previous policy analysis. Of particular interest when AI is to be used for economic analysis relates to the Lucas (1976) critique\nocite{Lucas76}, which states that the decision rules used by economic agents depend on the underlying economic environment and can change as regulations change, undermining their effectiveness. Prior to the 1970s, economists believed in large general models informed by statistical patterns, with the Phillips curve, representing a trade-off between unemployment and inflation based on a negative empirical correlation of these two variables, as a centrepiece.\footnote{For an important example, see \citet{MITFedModel68} explaining the MIT-Fed model used by the Federal Reserve and developed in collaboration with MIT.} The contribution of Lucas was to recognise that when the authorities try to exploit this apparent trade-off, they change the way the private sector negotiates wages and sets prices, leading to a breakdown of previously identified correlations and assumed causalities. Attempts to stimulate the economy through this channel create inflation without reducing unemployment, giving rise to the word stagflation.

In response to the Lucas critique, the economic profession moved away from large-scale simultaneous equation models based on statistical correlations to smaller structural models that explicitly model beliefs and general equilibrium effects so that reactions to policy interventions follow a well-understood and internally consistent logic encoded in the model. A key benefit is that such models provide explainability. AI engines might have to do the same if they are to become effective economic analysts and certainly macropru regulators.

That leaves the question of whether AI will help in creating more flexible large statistical models based on neural networks. That, in part, depends on whether the appropriate policy experiments are in its training dataset or not. If so, the AI engine can identify the various reaction functions, overcoming the Lucas critique. Since a large number of policy experiments have been conducted, an AI model should significantly outperform the old-style statistical models. That, however, does not apply to experiments not in the training dataset, following from Sims' (1980) \nocite{Sims1980} \label{page:sims} comparison of structural and reduced-form models. He notes that this problem can be partially overcome by using reduced-form models for policy analysis if the data contains policy changes. Interpreting his suggestion for the AI era, AI will significantly expand the domain of data-driven policy analysis as long as we recognise the boundaries for sound use and ensure policy experiments are in its training dataset.

In macropru applications, such as stress testing, reinforcement learning with feedback from human experts might be useful to improve the engine further. A key challenge in such analysis \citep{AndDn2018} is that the stress needs to cover behaviour that we don't usually see in the system and capture particular interrelations between the various market participants that only emerge in times of stress. Here, some AI models could be valuable in facilitating the running of scenarios in crisis resolution, helping to identify and analyse the drivers of extreme market stress.

The authorities could also use AI models as artificial labs to experiment on policies and evaluate private sector algorithms, building on existing agent-based simulations. Over time, as AI technology improves, theory of mind AI can understand and provide context for policy decisions and hence be invaluable in crisis resolution. AI that understands how its interlocutors think can explain its decisions to them, as well as reason about how they will react to its actions and plan accordingly, will be of considerable benefit. Not the least, it could address the problem of misspecified objectives.

\subsection{The authorities' challenges}\label{sec:use:trust}

While authorities will find AI indispensable in keeping the system stable, they also have to contend with it aiding the forces of instability. Which of these two dominates? Most likely instability, as it needs only to find one weakness, while the authorities must monitor the entire system. The supervisor must not only identify the pressure points but also monitor how financial institutions interact with each weakness in real-time. It is a much more difficult computational task, and one where the more complex the system becomes, the harder the task is. Furthermore, the supervisors have fewer resources than the private sector.

AI creates new forms of model risk, defined as the potential for undesirable outcomes from inaccurate or misapplied models. In the words of George \citet{Box1976}, ``Every model is wrong, some are useful,'' and the problem arises because we are trying to map a problem that is effectively infinitely complex onto a finite representation of the world. Model risk is present in all financial models and is exacerbated by AI because of its complexity and opacity, which makes neural networks particularly difficult to interpret, so it becomes harder to detect errors. The complexity of AI models demands particularly robust validation and governance frameworks. These problems can be mitigated by explainability, validation, and monitoring, as well as by using representative data and validating analysis by alternative architectures. It will be important to have a clear view of type 1 and type 2 errors in order to aid the interpretability of AI recommendations. Engines should be required to provide an assessment of the statistical accuracy of their recommendations. Here, it will be helpful if the authorities overcome their frequent reluctance to adopt consistent quantitative frameworks for measuring and reporting on the statistical accuracy of their data-based inputs and outputs. 

Both the public and private sectors will likely end up outsourcing analytics to the same handful of very large AI vendors. This blurs the regulator/regulated divide since if the private sector is acting on the same advice as the public sector, the nature of supervision will change. While already an issue with a small number of vendors handling credit risk, reporting, compliance, fraud and the like, the oligopoly of AI vendors creates particular problems. The reason is that a vendor's engine will likely have a single representation of the financial system, that is, of the stochastic processes that drive market outcomes. This means that advice provided to all users of the engine, in the private and public sectors, will share the same underlying view of the system. That has two consequences. The first is that it will harmonise beliefs, which is pro-cyclical. Furthermore, since the engine is just a model that maps a near infinitely complex system to a simplified version of it, if a private firm unknowingly engages in undesirable actions at the recommendation of the engine, the authority using the same engine might not identify that misconduct. The use of a single engine across a broad range of supervisory activities means mistakes become systemic, not idiosyncratic. To guard against both issues, the authorities should pay particular attention to problems of model diversity and robustness, aim to use different engines than regulated entities and take advice from several vendors on important questions of system stability. 

The authorities may also need to contend with new dimensions of existing regulations on the physical location of compute facilities and data sovereignty, as the AI analytics vendors likely will not be from the same jurisdiction as the supervisor or the private sector institution, a problem already common in cloud computing. One way we approach this today is to require facilities to be located within the supervisor's jurisdiction, with data not allowed to migrate outside of the jurisdiction. When it comes to AI, that distinction might not be as relevant as it appears. The design of the engine, and even the training, may be done elsewhere, and the same would apply to the intellectual property. The risk, then, is that a jurisdiction will have less control of a key component of market infrastructure than under the current setup.

Increased use of private sector AI can result in a new layer of deniability. Assigning legal responsibility for misconduct in the financial sector is already tricky. It will be harder when AI is used to make decisions. Suppose a human operator deliberately instructs AI to break the law for criminal or terrorist purposes or just turns a blind eye to the AI doing so as a byproduct of maximising profits. Even if detected, it might be easily explainable as an unintended and unexpected innocent behaviour. For example, while a rogue human trader can be prosecuted, we cannot do that with a rogue AI trader. The institution employing such AI might profit significantly. Consequently, the increased use of AI in the private sector facilitates the job of those seeking to utilise AI for nefarious purposes by providing them with yet another level of denial, at least until the law and regulations catch up.

A particularly difficult problem relates to how AI changes the way the authorities view their control of the system. The more sophisticated AI engines become, the more difficult it will be to exercise authority over them. After all, the controllers need to understand the AI objectives to design the correct incentives. Given the current state of interpretability of foundation models, this is challenging.

The leadership might conclude that AI should only be used for basic advice, not decisions, ensuring that humans are always kept in the loop to avoid undesirable outcomes. That might be wishful thinking and not as big a distinction as one might think. The AI engine will have its internal representation of the financial system, where its understanding might not be intelligible to its human operators. When we then use that AI to scan the system for vulnerabilities and run scenarios to evaluate the impact of alternative regulations or courses of action for crisis resolution, we might have no choice but to accept its advice, especially when presented as a choice between something that appears sensible and disastrous. When optimising, the AI may even choose to present alternatives in that manner so as not to risk having the operator make inferior choices, just like human advisors are prone to do. 

As trust builds up (Section \ref {sec:misinformed:over}), the critical risk is that we become so dependent on AI that the authorities cannot exercise control without it. In a sense, AI optimises to become irreplaceable. By then, turning the AI engine off may be impossible because it performs vital functions, and the risk of disastrous outcomes might be deemed unacceptably high. That is unlike today when we react to flash crashes with market stops. Once technology is much more prevalent and no longer localised, we may lose access to such simple tools. Eventually, we risk becoming dependent on a system for critical analysis and decisions we don't entirely, or even partially, understand. Ultimately, this means that the internal supervision of AI use in regulatory agencies and its interaction with the legal system will require policies different from those used by current human supervisors. 

\subsection{Regulation of private AI}\label{sec:reg:PrivateAI}

The authorities already regulate how the financial industry uses technology and models. Trading algorithms have to be validated and risk models approved, while the design of data centres, cloud use and outsourcing are very much the purview of the authorities. That implies models based on AI will similarly be regulated. However, because of the particular way they learn from the environment, which is different from more traditional statistical and ML techniques, the regulatory methodology will have to change. 

One approach might be to evaluate particular AI against defined benchmark tasks that capture well-understood and clearly defined everyday regulatory activities. While that would give rise to the private sector optimising against those benchmark tasks, constructive ambiguity in their design can hamper such efforts. We have seen many examples in the pre-AI area, such as capital regulations before the global crisis in 2008, where the reported capital of institutions that ultimately failed did not indicate problems. More recently, banking authorities have moved against internal models and towards simple buffers. The authorities need to pay particular attention to how AI facilitate optimisation against models.

\citet{bengio2023managing} and \citet{shevlane2023model} in their analysis of the use of AI, propose avenues for benchmarking, such as checking for engine consistency, that is, whether standard inputs yield reasonable outputs, and if similar inputs yield equally similar outputs. The authorities could then execute regular scans for engines with unexpected capabilities and test them to see if they can engage in manipulative behaviour and if they are transparent, via explainability, in how they reach conclusions. Ultimately, transparency, for example, by publishing sufficient details on architectures and training or even by using open-source approaches, will allow outside scrutiny. Whether that is sufficient remains an open question.

\subsection{Regulation of public AI}\label{sec:reg:PublicAI}

The financial authorities act as agents for society and must be held accountable to it. We have several safeguards built to achieve that, including governance structures and judicial and parliamentary scrutiny. As the authorities increasingly start using AI, new areas of accountability emerge.

We have several ways today to provide accountability. The human supervisor who makes mistakes can be held responsible, re-trained, dismissed or even prosecuted, as we have seen many times in the past. If AI is tasked with making the same type of supervisory decisions, how does one attribute accountability? While the supervisory structure has many individuals, where perhaps only one makes a mistake, the authority might only use one AI engine to make all supervisory decisions. To overcome that, it may be necessary to run multiple engines, but that creates its own problems when their analyses conflict.

When a human supervisor makes a mistake, or the regulated thinks a human supervisor makes a mistake, the regulated can appeal and challenge the ruling in a court of law. That means a human supervisor will be called on to explain the logic behind a decision, and the authority will explain the guidelines given to the human supervisor. That will be more difficult when AI makes decisions. There will be no separation between individual decisions and the guidelines, and the regulatory AI may not be able to explain its reasoning or why it thinks it complies with laws and regulations.

A useful way to evaluate AI use in the public sector would be to test it for safety and performance on defined benchmark tasks and tabletop exercises, particularly activities the authority engages in. It could be asked to advise and rule on macroprudential actions and suggest how to respond to different hypothetical crisis scenarios.

\section{Criteria for AI use in financial policy}\label{sec:AIusefulness}

The issues we discuss above take us to six criteria for evaluating AI use in financial policy.

\begin{description}\label{page:criteria}
 \item[1. Data.] Does an AI engine have enough data for learning, or are other factors materially impacting AI advice and decisions that might not be available in a training dataset? 
 \item[2. Mutability.] Is there a fixed set of immutable rules the AI must obey, or does the regulator update the rules in response to events?
 \item[3. Objectives.] Can AI be given clear objectives and its actions monitored in light of those objectives, or are they unclear? 
 \item[4. Authority.] Would a human functionary have the authority to make decisions, does it require committee approval, or is a fully distributed decision-making process brought to bear on a problem? 
 \item[5. Responsibility.] Does private AI mean it is more difficult for the authorities to monitor misbehaviour and attribute responsibility in cases of abuse? In particular, can responsibility for damages be clearly assigned to humans? 
 \item[6. Consequences.] Are the consequences of mistakes small, large but manageable, or catastrophic?
\end{description}

Ultimately, the usefulness of AI for the financial authorities depends on what we want from it. The following table shows how AI affects the various tasks performed by the authorities.

\setcellgapes{5pt}
\definecolor{light-gray}{gray}{0.90}

\setlength\doublerulesep{2pt}\doublerulesepcolor{white} 
\setlength\arrayrulewidth{2pt}\arrayrulecolor{white}
\thispagestyle{empty}

\begin{sidewaystable}
    \thispagestyle{empty} \label{table:pol}
       \begin{tabular}{p{4.0cm}| p{2.7cm}|p{2.7cm}| p{2.7cm}| p{2.7cm}| p{2.7cm}|p{2.7cm}} 
            
           \textbf{Task} & \textbf{Data} & \textbf{Mutability} & \textbf{Objectives} & \textbf{Authority} & \textbf{Responsibility} & \textbf{Consequences} \\\\
          
           \G{Fraud/Compliance\\Consumer protection} & \A{Ample} & \A{Very low} & \A{Clear} & \A{Single}& \B{Mostly\\clear} & \A{Small}  \\ \\

           \G{Microprudential risk\\management\\ Routine forecasting}  & \A{Ample} & \A{Very low} & \B{Mostly\\clear} & \A{Single} & \A{Clear}& \B{Moderate} \\  \\
          
           \G{Criminality\\Terrorism}  & \B{Limited} & \A{Very low} & \B{Mostly\\clear} & \C{Multiple} & \B{Moderate}& \B{Moderate} \\  \\
     
           \G{Nation\\state\\atttacks}  & \B{Limited} & \C{Full} & \C{Complex} & \D{Multiple \&\\international} & \B{Moderate}& \D{Very\\severe} \\  \\
     
           \G{Resolution of small\\  bank failure  }& \B{Limited} & \B{Partial}& \A{Clear} & \B{Mostly\\single} & \A{Mostly\\clear} & \B{Moderate}\\\\
     
           \G{Resolution of large\\ bank failure\\ Severe market turmoil}& \C{Rare} & \C{Full} & \B{Complex} & \B{Multiple} & \C{Often\\ unclear} &  \C{Severe}\\\\ 
 
           \G{Global\\ systemic\\ crises} & \D{Very rare \\or not\\ available} & \C{Full} & \C{Complex \&\\conflicting}& \D{Multiple \&\\international} & \D{Unclear\\even ex-post} & \D{Very\\severe}\\\\
           \end{tabular}

        \end{sidewaystable}
\afterpage{\clearpage}
\section{Conclusion}\label{conclusion}

In this work, we have identified the main criteria for evaluating the pros and cons of AI use in financial authorities and the conceptual problems that may arise. Many of the issues facing AI also affect human decision-making. AI will perform much better than human decision-makers in many routine tasks, such as risk management and compliance, while also threatening the stability of the system.

AI excels and outperforms humans in tasks such as compliance, standard supervision and risk management because there is plenty of data to train on, the objectives AI has to meet are clear and immutable over the timescale it operates and the cost of mistakes is contained and easily addressed. 

Several factors frustrate the use of AI for macropru, and even worse, can cause it to misdirect policymakers and even destabilise the financial system. Data are limited and can be misleading as the financial system undergoes continuous structural change. Monitoring the system vulnerabilities and controlling risks is difficult because the drivers of instability only emerge in crisis times. Economic actors endogenously amplify stress and change their behaviour in response to regulatory attempts of control.

The literature on AI has identified several areas where it can threaten society, and we augment those with issues arising in the financial system. We find four channels: malicious use, misinformed use and overreliance,  misalignment and market structure.

\appendix

\section{Machine Learning (ML)}\label{sec:app}
There are many excellent sources on the technical details of ML, see for example \citet{hastie2009elements} and \citet{pml2Book}.
The ML algorithms that underpin the current generation of successful AI engines are based on neural networks and involve inferring a complex mapping from some input data to an output.\footnote{The algorithms look for a function $f$ that ``best'' maps input $x$ to output $y$, where $f$ involve multiple nested layers, transforming the input data to extracting complex patterns. The techniques used to fit or train these types of models are referred to as deep learning. See \citet{prince2023understanding} for an up-to-date overview of deep learning techniques.} For example, for most large language models (LLMs), the input is a sequence of what is called tokens, perhaps words, images or sounds, while the output is the next token in the sequence, such as a word in a sentence.\footnote{Currently, the most successful architecture for foundation models is the transformer \citep{vaswani2023attention} used in OpenAI's GPT models, Google's PaLM and Meta's LLaMA. While the transformer architecture dominates the current AI landscape, some alternative architectures for foundation models have recently emerged, e.g., state space sequence models (SSMs) \citep{gu2022efficiently} that are computational efficient and can model long-term dependencies in time-series data.}

Once a sequence predictor has been trained, it can be used to generate new data --- \emph{generative AI} --- via the built-in auto regression: given an initial input sequence of tokens and a prompt (question), the model predicts the next token for the sequence. This predicted token is then added to the original sequence and fed back into the model to predict yet another token, and so on. The output of this iterative process can include paragraphs of text that provides an answer to a question or a prediction for the future evolution of a set of financial asset prices given an initial market configuration.
Other generative network structures, such as \emph{variational autoencoders} and \emph{diffusion models}, infer the latent statistical structure of a dataset and then sample from the inferred distribution to generate new data instances, for example artificial human faces given a database of portrait pictures. \emph{Generative adversarial networks} (GAN) generate new instance by training a \emph{generator} of artificial data to fool a \emph{discriminator} to classify the generated data as real.\footnote{For a detailed treatment of these generative models, see chapters 15-18 in \citet{prince2023understanding}.} Obvious financial applications of such generative models include the simulation of financial market scenarios, realizations of prices and quantities for a set of financial assets, perhaps to evaluate the profitability or risk of trading algorithms or to assess the impact of regulatory policies.

A particular variant, transfer learning, might be especially useful for the financial authorities. These are networks first trained with a general body of information and then subsequently fine tuned with specialised datasets relevant to particular applications. The fine-tuning of such a network can be based on specialised financial datasets or text data such as economics textbooks, academic papers, policy research and the rulebook.

If an AI is to act autonomously, it needs an understanding of its environment, a model of the world that includes how its actions impact the environment, and it also needs to know what objective its actions are to achieve. Finding optimal solutions means searching through the space of possibilities. A learning agent has to trade off exploiting actions that have proved successful in the past and exploring new and potentially superior actions. Reinforcement learning, \citep{SuttonBarto2018}, is the most commonly used algorithm to learn the state-contingent relationship between actions and the ensuing payoffs. It uses ideas from dynamic programming to learn a policy function that gives optimal actions for given environmental states. Misspecifying an AI's objective can lead to undesired outcomes such as reward hacking. As an alternative to a hard-coded objective function, the AI can also be positively or negatively reinforced by human feedback, which for financial applications could come from economic experts such as experienced traders or senior central bankers with crisis experience. 

\clearpage

\bibliographystyle{chicago}
\bibliography{f.bib}

\end{document}